# USING CONTENT FEATURES TO ENHANCE THE PERFORMANCE OF USER-BASED COLLABORATIVE FILTERING

Niloofar Rastin and Mansoor Zolghadri Jahromi

Department of Computer Science and Engineering, Shiraz University, Shiraz, Iran

## ABSTRACT

*Content-based and collaborative filtering methods are the most successful solutions in recommender systems. Content-based method is based on item's attributes. This method checks the features of user's favourite items and then proposes the items which have the most similar characteristics with those items. Collaborative filtering method is based on the determination of similar items or similar users, which are called item-based and user-based collaborative filtering, respectively.In this paper we propose a hybrid method that integrates collaborative filtering and content-based methods. The proposed method can be viewed as user-based Collaborative filtering technique. However to find users with similar taste with active user, we used content features of the item under investigation to put more emphasis on user's rating for similar items. In other words two users are similar if their ratings are similar on items that have similar context. This is achieved by assigning a weight to each rating when calculating the similarity of two users.We used movielens data set to access the performance of the proposed method in comparison with basic user-based collaborative filtering and other popular methods.*

## 1. INTRODUCTION

With sudden increase in volume of information, the need for a tool to assist users find the information they seek became apparent. In this context, recommender systems have played a significant role. These systems are software tools and techniques that introduce the items according to user needs. Items can be movies, music, web page, and so on.

Two key strategies of recommender systems include collaborative filtering and content-based methods. Of course, that there is a third method of integrating the two mentioned methods which leads tohybrid recommender system.

The workflow of Content-based system is that at first a profile of user's interests is made on the basis of the way that user has rated various items. Then, based on the compliance of item's features with the profile made of the user, the user is presented with some suggestions.

Collaborative filtering recommender systems are divided in two main groups of memory-based and model-based approaches [1]. In memory-based collaborative filtering the Matrix of user-item is checked and used directly for prediction. In this matrix, the scores which the users have given to the various items are shown. The advantage of this method is that all the information is available anytime. But with enlarging the matrix and growing number of users and items, search space, required memory and computation time will increase. In memory-based approach, the prediction is performed based on users or items [1]. Prediction based on users, is done by the evaluation of scores which the same users have allocated to the desired item. Similar users are those whose scoring pattern is similar to the active user [2]. Prediction based on items, is done by the evaluation of scores which the active user have allocated to the similar items. The items are similar if many users similarly voted for them [3]. Unlike memory-based collaborative filtering in which stored scores are directly used in predictions, model-based collaborative filtering uses available scores to learn a prediction model. This model based on available data is taught as offline and then is used as online in order to predict the scores that would denote to the new items [4]. So it is faster in comparison to memory-based collaborative filtering. This model can be a machine learning algorithm such as Bayesian network, clustering, and rule-based approaches [3].

Both Content-based and collaborative filtering methods have some strong and weak points. According to the obtained understanding of the problem, we should decide which method is more appropriate to use. For example, if the number of items is limited, the chance of collaborative filtering method to be successful is low and if the descriptive information about the items is not available, content-based method cannot be used.

This work relies on user-based collaborative filtering method. In this method, similar users are identified based on their ratings of the items, then rating the items that have never been seen is predicted, and finally the items that have high ratings are recommended to the user. In this method, all the items have equal effect on determining the degree of similarity between the users. But in fact to predict the target item rating, similarity of user ratings to items similar to the target item, are more important than other items.

In this paper, a content-based method is provided to weight the items in collaborative filtering systems. Because the dataset used in this paper is MovieLens and related to the movie, the meaning of items is the movies available in dataset.In order to use content-based method for weighting the items, the characteristics of genres, directors and actors of each movie have been studied. The genre of each movie is the specific category of that. For example, if the genre of a movie is comedy-drama, it means that the movie belongs to two categories of comedy and drama. In mentioned dataset, the information of each movie's genres is available. In this dataset, there are 19 general genres and each movie has a minimum of 1 and maximum of 3 genres. In addition to the features of the genres of each movie, other data been also used such as the features of movie's directors and actors. These features do not exist in the dataset and are extracted from the Linked Open Data (LOD) datasets, such as DBpedia. It is worth noting that the use of additional data is to weight more accurately the items by content-based method and consequently enhancing the accuracy of prediction in collaborative filtering systems.

The remainder of the paper is structured as follows: section 2 gives an overview of the proposed architecture. The implementation of the case study is described in section 3. Section 4 presents the experiments and the results obtained. In the last section, the paper provides a short conclusion and overview of future work.

## 2. RELATED WORKS

A variety of strategies are presented to determine the effectiveness of the items in collaborative filtering that are briefly described in the following.

[5], is the criterion of Inverse Document Frequency, which is a known criterion in information retrieval, is used to weight the items in collaborative filtering system. The main idea of this method is called the inverse user frequency. The items that are popular among users in general

can't accurately reflect the interests of a user. Therefore, these items should be allocated less weight than other items.

[6], has presented an idea similar to [5], and in this idea, the variance strategy is employed in order to lose the weight of popular items. So that, the Items which have more scores variances, may obtain more weight.

[7], has presented an approach based on information theory. In this approach using the criterion of mutual information and entropy, the degree of dependence between the target item and other items is determined and based on that, the weight of items are allocated to them.

[8], proposes an automatic weighting method that uses the idea of model-based systems. This method by maximizing the average similarity between users, allocate weights to the items. So that makes the user closer and more similar to those with similar tastes and makes him more distinctive from who has different taste.

Because of the conflicting reported results, [9] has offered a comparison between different methods of weighting the items, and also three ways to filter the items based on the weight allocated to them has been introduced.

[10], has solved the problem of similarity of items and sparsity of collaborative filtering systems, by the local and general similarity of users. In a way that local similarity between users is calculated by decreasing the influence of the popular items among the population. It is done by considering each item as a random variable of the Laplace distribution.

[11], has proposed a new mechanism for weighting the items and overcome the problem of sparsity. This approach is based on Latent Semantic Analysis and the method of Singular Value Decomposition.

[12], has the duty of checking the cold start problem during weighting the items. Weighting the items based on reducing the impact of popular items is done by the use of two methods including inverse user frequency and the linear weighting.

Most methods which were proposed for allocating weights to the items, have used the statistical data of the items which means the scores allocated to them, While it is possible to use the content of items in order to determine the similarities of them and weighting them. Hence, in this paper, a content-based method is used for weighting the items that are described in more detail in the following.

## 3. PROPOSED METHOD

The proposed method consists of the following three distinct stages:

1 - Pre-processing
2 - Allocating weights to items based on content-based method
3- The use of weight allocated to the items in two phases including neighbour selection and prediction in collaborative filtering method

Steps mentioned above are described in detail in the following descriptions.

## 3.1. Pre-processing

As it was previously noted, in order to use the features of actors and directors of each movie in content-based method, we need to extract those features from DB Pedia dataset.DB Pedia extracts and organizes the existing information contained in Wiki Pedia and make them available. In order to use the information related to the movie, a query based on the movie title was designed in SPARQL language and by Post URL method is sent to the DB Pedia server. A sample of designed application is presented in the following.

```
SELECT ?film_title  ?star_name  ?nameDirector {
    {
      SELECT DISTINCT  ?movies  ?film_title
        WHERE {
            ?movies  rdf:type  <http://dbpedia.org/ontology/Film>;
rdfs:label  ?film_title.
        }
    }.
 ?movies dbpedia-owl:starring ?star;
 dbpedia-owl:director ?director.
 ?director foaf:name ?nameDirector.
 ?star foaf:name ?star_name.

 FILTER ((str(?film_title) IN ("Film Name"))
&&(LANGMATCHES(LANG(?film_title),"en")))
 }
ORDER BY ?film_title
```

This Server sends related information of the desired movie in XML format. This is done for all movies in the MovieLens dataset. It should be noted that the extracted information for each movie includes the names of all directors, and just the names of the starring actors. Information about the movies that is not available in DBPedia is extracted manually from Wikipedia website, and if it was not available, it would be extracted from IMDB website. In cases where information is extracted from the IMDB website, consistently the names of seven top actors in term of their credits have been selected as stars.

## 3.2. Weighting The Items

After gathering information about the genres, directors and actors of movies in dataset in the previous step, we now turn to weighting them based on content-based method. Be given that the information about each movie, represent the profile of that movie. So the weight of each movie is determined based on its similarity to the profile of target movie. The similarity is measured by the cosine-based similarity measure. This measure in general, receives two vectors as input and measures their similarity based on the angle between them. To determine the degree of similarity between the two movies based on this measure, the information contained in the profiles of both items should be shown in vector format with same length. To this end, for each feature of genre, director and actor belonging to the profiles of either two comparing movies, a component is considered in a vector. Then the values 0 and 1 are used to indicate the presence or absence of the features in each movie.

Suppose that $G_i$ represents $i$-th genre, $D_i$ represents $i$-th director and $A_i$ represents $i$-th actor in a movie. Consider movies $M$ and $T$ with following features:

$M = \{G1, G2, D1, D2, A1, A2, A3\}$
$T = \{G1, G2, G3, D3, A2, A3\}$

The conjunction of the feature set of these two movies is:
$M \cup T = \{G1, G2, G3, D1, D2, D3, A1, A2, A3\}$

Thus, a built vector for each movie is as following:
$M = \{1,1,0,1,1,0,1,1,1\}$
$T = \{1,1,1,0,0,1,0,1,1\}$

According to what was described in the pre-processing stage, the number of actors extracted for each movie is different. So when you compare any movie with the target movie, common actors between them are just considered. So in above example, actor $A1$ will be deleted from $T$ and $M$ vectors:

$M \cup T = \{G1, G2, G3, D1, D2, D3, A2, A3\}$
$M = \{1,1,0,1,1,0,1,1\}$
$T = \{1,1,1,0,0,1,1,1\}$

Now that we could show the movie's profile in the form of vectors, we use cosine-based similarity measure to determine the similarity between two vectors in the following way:

$$w_{M,T} = \cos(\vec{M}, \vec{T}) = \frac{M.T}{\parallel M \parallel_2 \times \parallel T \parallel_2} = \frac{\sum_i M_i T_i}{\sqrt{\sum_i M_i^2}\sqrt{\sum_i T_i^2}} \qquad (1)$$

$M_i$ Represents $i$-th component of vector $M$ and $T_i$ represents $i$-th component of vector $T$. Note that the result of this fraction is a value between 0 and 1. 1 means perfect similarity and 0 means perfect dissimilarity.

Since the components of two compared vectors are 0 and 1, Calculated value in numerator of above fraction is equal to the number of shared ones (1s) and in the other words it is equal to the numbers of common features between two movies. So, zero weight is considered for the items that have not any common feature with target item. On the other hand, the basis of collaborative filtering is calculating the similarity between users that some of these users have a limited number of rated items. Thus, there are items that have not got enough scores yet. This makes the problem of sparsity of user-Item matrix. In this case, measuring the similarity is done on limited number of expressed scores that cannot be trustworthy. Now, this problem will be exacerbated by zero consideration of items that have not common features with the target item. To avoid this problem, a smaller weight is considered for these items in comparison with other items. Thus, the weight is applied to the items as following:

$$w_{M,T} = \frac{1 + \sum_{i=1}^{n} M_i \times T_i}{\sqrt{\sum_{i=1}^{n} M_i^2} \times \sqrt{\sum_{i=1}^{n} T_i^2}} \qquad if\ k \geq 1$$

$$w_{M,T} = \frac{1}{\sqrt{\sum_{i=1}^{n} MV_i^2} \times \sqrt{\sum_{i=1}^{n} T_i^2}} \qquad otherwise \qquad (2)$$

$K$ specifies the number of shared features between two items (the number of ones between two vectors). $MV$ Represents a movie that has the greatest number of features (a vector with the largest number one).

### 3.3. Neighbour Selection

According to what was explained, result obtained from the previous stage is the weight of each item based on its similarity to the target item. To predict and provide suggestions by collaborative filtering method, we must first find the most similar users to active user and consider them as a set of the neighbours of active user. Active User is a user whom the goal of the prediction of target item's score is for him. In order to create the set of neighbours, only the members who voted for target item have been studied. Standard Pearson correlation Criteria is used for evaluating the dependencies between scoring patterns of users and active user, and can be calculated as follows:

$$PC(a,u) = \frac{\sum_i (r_{a,i} - \bar{r}_a)(r_{u,i} - \bar{r}_u)}{\sqrt{\sum_i (r_{a,i} - \bar{r}_a)^2 \sum_i (r_{u,i} - \bar{r}_u)^2}} \qquad (3)$$

$\bar{r}_a$ and $\bar{r}_u$ are the averages of total scores that the users $a$ and $u$ have attributed to the items. $r_{a,I}$ and $r_{u,I}$ are the scores that users $a$ and $u$ have attributed to the $i$-th item respectively.

At this point, to measure the dependence between user rating patterns with the active user, Weighted Pearson Correlation is used. The only difference of this criterion with the criterion of standard Pearson Correlation is that at the time of comparing the way of scoring of two users to each item, also the weight of the item will be involved. The weight, determines the importance of similar performance of both users in rating this item. Weighted Pearson Correlation is as following:

$$WPC(a,u,j) = \frac{\sum_i (w_{j,i}(r_{a,i} - \bar{r}_a))(w_{j,i}(r_{u,i} - \bar{r}_u))}{\sqrt{\sum_i (w_{j,i}(r_{a,i} - \bar{r}_a))^2 \sum_i (w_{j,i}(r_{u,i} - \bar{r}_u))^2}} \qquad (4)$$

$j$ represents the target item that predicting it's score is desired. $W_{j,I}$ represents the weight of $i$-th item. In other words, it is the amount of similarity between $i$-th item and the target item.

Additionally, by declining the correlations based on limited number of co-rated items in combination with significance weighting, we can have a great increase of prediction accuracy. Assuming that $x$ is the number of items that users $a$ and $u$ have commonly voted, finally the similarity of two users is obtained by using the equation 5.

$$sim(a,u) = WPC(a,u,j).CF \qquad (5)$$

$CF = 1 \qquad$ *if* $x > 50$

$CF = \frac{x}{50} \qquad$ *otherwise*

After that the similarity of all users with active user is distinguished, it is the time of choosing the set of neighbours. Therefore users would be arranged in descending order based on their resemblance. We have to select the best among these users. It has been shown that it is better to select a fixed number of their best (potentially in a range from 20 to 60) than to use a similarity weight threshold [13].

## 3.4. Prediction

The result of the previous stage is the set of active user's neighbours. At this stage, by the use of the scores allocated to target item by set of the neighbours, the score of target item can be predicted. For this purpose, the equation 6, which is typically used in user-based collaborative filtering [14], is applied.

$$\bar{r}_{a,t} = \bar{r}_a + \frac{\sum_{u \in N_{t(a)}} (r_{u,t} - \bar{r}_u).sim(a,u)}{\sum_{u \in N_{t(a)}} sim(a,u)}$$  (6)

$N_{t(a)}$ is set of active user's neighbours. $\bar{r}_a$ and $\bar{r}_u$ are the average of total scores that the users $a$ and $u$ have attributed to the items respectively.

# 4. EXPERIMENTAL RESULTS

The evaluation settings are provided in this section and also the experimental results for the model are presented. We aim to evaluate the method as an alternative to user-based collaborative filtering recommender systems, which can complement and improve the available methods.
Proposed method has been studied and tested on vogue and popular Movie Lens dataset that is related to the recommender site of the movie. This dataset is composed of 1,000,209 scores (on a scale from 1 to 5) that are allocated to 6,040 movies, by 3,959 users. To test the proposed method 5-fold cross-validation is used. This means that the scores given to each movie is almost equally divided into 5 parts, then a portion of 20% would be for testing and the remaining part would be for training. So that by means of 80% of scores, 20% of the remaining scores is predicted. Totally, the test set contains nearly 192,710 scores and testing set would have almost 807,499 scores.

An assessment criterion of recommender system is selected based on their duty. At this point, since the aim is evaluation of the ability of recommender system in prediction of unobserved item's rating, mean absolute error is applied to evaluate the proposed method. Mean absolute error is based on the accuracy and measure the distance between predicted and actual scores. This criterion is calculated as following:

$$MAE = \frac{1}{n} \sum_{i=1}^{n} (r_i - \hat{r}_i)$$  (7)

$r_i$ is the actual score, $\hat{r}_i$ is the predicted score and $n$ is the number of predicted scores.

The mean absolute error of the proposed method is compared with standard collaborative filtering method which was presented as baseline in[14]. It should be noted that the equations 4 and 5 were used to select the set of neighbours and the equation 6 is used for prediction. Results of mean absolute error in basic method (PC) and the proposed method (WPC) are shown in Table 1 and it's diagram is shown in Figure 1. In this diagram, the vertical axis shows the mean absolute error and the horizontal axis represents the number of neighbours.

Finally, The results obtained from the proposed method with the results obtained from prediction-based method [15] and also the results of the core-based method [16] which is one of the strategies proposed in this area, are compared with each other. It is noteworthy that the core-based method has achieved significant results on different datasets and is one of the state-of-art technologies for the MovieLens dataset [15]. Table 2, summarizes the best results obtained by the proposed method and two other mentioned methods.

Table 1. Comparing the mean absolute error
of basic method (PC) and proposed method (WPC)

| Number of Neighbours | PC | WPC |
|---|---|---|
| 5 | 0.7463 | 0.7256 |
| 10 | 0.7179 | 0.6967 |
| 20 | 0.7076 | 0.6834 |
| 30 | 0.7055 | 0.6804 |
| 50 | 0.7046 | 0.6793 |

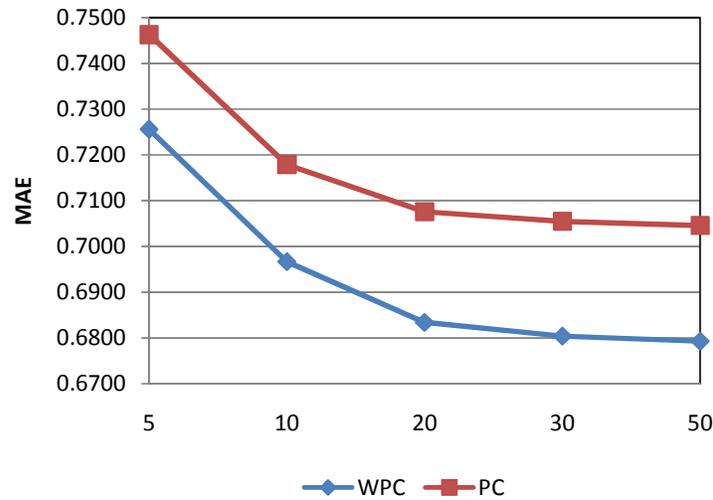

Figure 1. Comparing the mean absolute error
Of basic method (PC) and proposed method (WPC)

Table 2. Comparing the mean absolute error of
proposed method (WPC) with methods [15] and [16]

| WPC | [15] | [16] |
|---|---|---|
| 0.6793 | 0.6798 | 0.685 |

As shown in Table 2, the results obtained from the proposed method are almost equal with the results obtained from the methods [11] and [12], and represents the quality of the proposed method in this paper.

## 5. CONCLUSIONS

In this paper, a user-based collaborative filtering recommender system is presented, which is equipped with dynamic allocation mechanism to weigh the items. Weighting of the items aimed at improving the quality of the active user's set of neighbours and follows with enhancing the accuracy of predicting the scores of the items in collaborative filtering system. Weight of each item is specified based on calculating it's similarity with the target items by content-based method.

In future works, ontology can be applied to improve content-based method in order to have more precise weight for items and increase the accuracy of collaborative filtering method.

**Authors**

Niloofar Rastin is the MSc student in artificial intelligence of Computer Science and Engineering Department of Shiraz University. As one of her research interests, she focuses on machine learning especially recommender systems domain. Also, she is co-author of several articles in machine translation and evolutionary algorithms area.

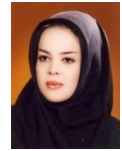

Mansoor Zolghadri Jahromi received his B.Sc. degree in Mechanical Engineering from Shiraz University (Shiraz, Iran) in September 1979, M.Sc. and Ph.D. degrees in Instrumentation and Control from the University of Bradford (Bradford, U.K) in 1982 and 1988, respectively. He joined the department of Computer Science and Engineering, Shiraz University in 1988, where he is currently working as a Professor. He is the author or co-author of more than 50 papers published in various international journals and conference proceedings. His current research interests include pattern recognition, fuzzy systems, information retrieval and web search

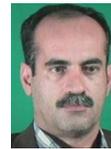